\documentclass[10pt]{article}
\usepackage{amsmath}
\usepackage{amssymb}
\usepackage{graphicx}
\usepackage{cite}
\usepackage{color}

\usepackage[labelfont=bf,labelsep=period,justification=raggedright]{caption}

\makeatletter
\renewcommand{\@biblabel}[1]{\quad#1.}
\newcommand{\epsWidth}{2in}

\makeatother

\date{}

\def\exp{\text{Exp\,}}

\def\log{\text{Log\,}}

\begin{document}
\begin{flushleft}
{\Large
\textbf{What does the Allen Gene Expression Atlas tell us about mouse brain evolution?}
}
\\

Swagatam Mukhopadhyay$^{1,\ast}$, Pascal Grange$^{1}$, Anirvan M. Sengupta$^{2}$ and Partha P. Mitra$^{1}$
\\
\bf{1} Cold Spring Harbor Laboratory, 1 Bungtown Road, Cold Spring Harbor, NY 11734
\\
\bf{2} BioMaPS and Dept. of Physics, Rutgers University, 174 Frelinghuysen Rd, Piscataway, NJ 08854
\\
$\ast$ E-mail: smukhopa@cshl.edu
\end{flushleft}

\section*{Abstract}
We use the Allen Gene Expression Atlas (AGEA) and the OMA ortholog dataset to investigate the evolution of mouse-brain neuroanatomy from the standpoint of the molecular evolution of brain-specific genes. For each such gene, using the phylogenetic tree for all fully sequenced species and the presence of orthologs of the gene in these species, we construct and assign a discrete measure of evolutionary age. The gene expression profile of all gene of similar age, relative to the average gene expression profile, distinguish regions of the brain that are over-represented in the corresponding evolutionary timescale. We argue that the conclusions one can draw on evolution of twelve major brain regions from such a molecular level analysis supplements existing knowledge of mouse brain evolution and introduces new quantitative tools, especially for comparative studies, when AGEA-like data sets for other species become available. Using the functional role of the genes representational of a certain evolutionary timescale and brain region we compare and contrast, wherever possible, our observations with existing knowledge in evolutionary neuroanatomy.  

%

\section*{Introduction} 

Investigations in brain evolution have traditionally used methods ranging from comparative neuroanatomy and paleontology, and an extensive literature exists on the subject~\cite{StriedterBook2005, ButlerBook2005}. We may be at the crossroads of a new era where the availability of gene expression data across the whole brain for multiple species will provide new quantitative tools in studying brain evolution. We propose a method whereby the molecular evolution of the mouse brain-specific genes, couple with the high resolution map of their brain-wide expression in the Allen Gene Expression Atlas (AGEA), can be used to draw conclusions on the evolution of the brain. We create maps for subsets of AGEA genes of mouse brain grouped by their approximate ``evolutionary age'' which is obtained by the analysis of orthologs of genes across all currently fully-sequenced species. These maps are a new resource in studying the evolution of major mouse brain regions, deduced entirely from molecular evolution of genes.  

The AGEA~\cite{Lein2007, Jones2009} maps the expression level of more than 20,000 genes in the mouse brain determined at high-resolution (200 $\mu{\mathrm{m}}$ voxel size) using \emph{in situ} hybridization of mRNA in a high-throughput manner. Such a high resolution dataset is unprecedented in any species, allowing us to formulate and address new questions about brain evolution using gene expression. Past work by other groups have employed various gene-expression datasets from multiple species for comparative studies of evolution in primates (including humans), where transcriptome of organs as a whole (including brain) was considered~\cite{Preuss2004, Khaitovich2006a, Vallender2008, Brawand2011}. Gene expression studies on the comparative neuroanatomy of avian and mammalian brains have generated a new set of hypothesis on brain-region homologies, for a review see Ref.~\cite{Jarvis2005}. For avian brain, gene-expression studies have focused on learned vocalization~\cite{Wada2004}. Currently, AGEA-like high-resolution gene-expression datasets is unavailable in any species other than mouse, however, creating a framework for systematic comparative study of brain evolution across species is of great interest. The methods we report here can be easily extended when such datasets do become available. In the current work, we restrict our attention to studying the evolutionary age of mouse brain-specific genes and their localization in brain regions to investigate the evolution of those brain regions. Such an investigation supplements traditional approaches in brain evolution. The question of whether signatures of molecular evolution of genes are informative in the evolution of brain regions is an important one especially because brain regions have been traditionally defined by considerations of function and morphology.

Gene-expression dataset have been used to study comparative brain evolution of brain-regions in humans. Ruppin group~\cite{Tuller2008} focused on the \textit{evolutionary rates} of genes expressed in twenty one different human brain regions using human and mouse brain tissue transcriptome~\cite{Su2002, Su2004}. Amongst other observations, the study highlighted the low evolutionary rate of genes over-expressed in the cortical brain region which are more recent in evolutionary time in comparison to the noncortical brain regions. They also observed that brain-specific genes have much lower median evolutionary rate compared to the rest
and genes that are more brain region-specific in their expression enjoy higher evolutionary rates. In contrast to their approach, we focus on evolutionary ages of mouse genes and not their rates. 

A closer comparison to the spirit of our method is the work by Grant group~\cite{Emes2008, Ryan2009}, which focused on the evolution of synapses used 19 species (8 mammals, 6 additional chordates and 5 other eukaryotes) and orthologs of a total of 651 mouse genes (corresponding to postsynaptic proteins) were identified by data-mining Ensembl Compara~\cite{Hubbard2009} and protein-BLAST~\cite{Altschul1997}. 
The work established that the core components of the synapse originated in unicellular eukaryotes, where the genes were involved in inter-cellular signaling and response to environmental stress. The post-synaptic genes most recent in origin are typically enriched in upstream signaling and structural components, and also contributed most to the variations of gene expression profile across the mouse brain, where the expression diversity was established by a hybrid of protein and mRNA data~\cite{Zapala2005}. Immunohistochemistry of mouse brain sections using antibodies to 43 different synaptic proteins was employed and mRNA were examined using \emph{in situ} hybridization maps~\cite{Magdaleno2006}. We use a similar ortholog search to ascertain the evolutionary ages of genes, but we consider a large subset of brain-specific genes.

In contrast to previous studies, we present brain-wide maps of `evolutionary ages' using more than three thousand genes and an up-to-date ortholog dataset for all 108 Eukaryotes sequenced. We harness the full potential of the AGEA for the purpose, and perform bootstrap analysis to ensure that the patterns of 'age'-grouped gene expression we report are statistically significant. The Allen Reference Atlas~\cite{DongBook2008} is the common three-dimensional atlas to which all the gene-expression data were registered, allowing us to study the signatures of gene evolution on major neuroanatomical regions with an unprecedented resolution. We also harness the statistical tools developed in determining the localization of gene-expression in AGEA---thereby reporting the most significant genes responsible in localized expression classified by a certain evolutionary timescale.

\section*{Results}

Our results are based on the analysis of orthologs across 108 Eukaryotic species and more than three thousand brain-specific mouse genes for which brain-wide gene-expression data was generated by the Allen Institute~\cite{Lein2007}. We computed correlation between the sagittal and coronal gene-expression datasets for 4104 genes and retained 3,041 genes that had the highest correlation across the two datasets. We use the OMA dataset to determine the orthologs of these genes for all the 108 species. The phylogenetic tree of these species are known, see for example Refs.~\cite{Hulsen2006, Hulsen2009}. We reason that the evolutionary age of a gene can be robustly mapped to a discrete score determined by considering the clade on the phylogenetic tree where almost all the orthologs for the gene appear. For example, if a mouse gene has orthologs only in vertebrates and not in any non-vertebrate chordate species, then the gene is a `chordate gene', i.e. it is at least as old as chordates. This discrete score, as opposed to a direct sequence-based-evolutionary time of divergence, is far less sensitive to uncertainties and noise (see Methods). One of the main challenges of any such study is the limited number of species fully sequenced, and the under-representation of certain clades in these species. For instance, there is only one amphibian (\emph{Xenopus tropicalis}) in the list. We strike a balance between probing clades of interest where major events in brain evolution is known to occur, and how well represented the clade is in order to draw statistically meaningful conclusions. Gene loss events can lead to correlated disappearance of orthologs in a subset of the clade leading to erroneous conclusion about the first appearance of the the gene on the tree. This problem is somewhat alleviated by including a fair number of genes in any age group (see Methods). 

We chose the following ordering of clades with respect to progressive refinement into subsets leading up to Rodentia. The last common ancestor the species within a clade appearing earlier in this ordering is older compared than the last common ancestor of a clade that appears later. 
\newline 
All $\supset$ \textbf{Eukaryota} $\supset$ \textbf{Metazoa} $\supset$ \textbf{Coelomata}  $\supset$ \textbf{Chordata} $\supset$ \textbf{Vertebrata} $\supset$ \textbf{Sarcopterygii} $\supset$ \textbf{Mammalia} $\supset$ \textbf{Eutheria} $\supset$ \textbf{Euarchontoglires}  $\supset$ \textbf{Rodentia} $\supset$ \textbf{Mouse} 


The result of the analysis of orthologs, as elaborated in Methods, is the assignment of a parameter $T \in [1,\dots, 12]$ to each gene $g$, corresponding to when $g$ first `appeared' in the above discrete ordering of clades. For an `old' gene $T$ is small, $T$ is large for a `new' gene. We denote by $G_T$ the set of genes appearing at $T$ and the set of all genes by $G$. 

In previous work involving two of the authors~\cite{Grange2011}, we have discussed the \emph{gene expression energies} $E(v,g)$ representing the AGEA data, where $v$ is the voxel index and $g$ is the gene index and $v$ is the voxel index. Briefly, (a) \emph{in situ} hybridization directly measures mRNA levels corresponding a particular gene $g$ in imaged section of the brain, denoted by gray-scale intensity $I(p,g)$ for pixel $p$, (b) a binary mask $M(p,g)$ is constructed to select for cell-shaped objects, (c) the following weighted sum of all pixels $p$ intersection a voxel $v$ defines $E(v,g)$---
\begin{equation}
E(v,g) = \frac{1}{N_p} \sum_{p\in v} M(p,g)\, I(p,g)
\end{equation}
where $N_p$ is the number of pixels contributing to a voxel. 

\begin{figure}
\begin{center}
\includegraphics[width=5in]{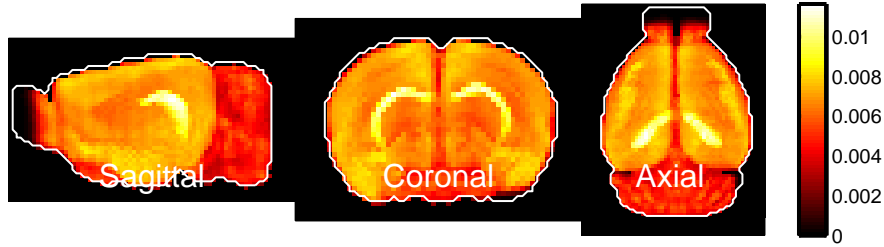} 
\end{center}
\caption{Maximal-intensity projection of the average of the set of 3041 genes in the Allen Gene Expression Atlas that have highest correlation between sagittal and coronal data.}
\label{fig:MIproj} 
\end{figure}

The AGEA also provides parcellations of the brain at various degrees of coarseness. One of them, which will be referred to as {\ttfamily{`Big12'}} defines 12 different regions of the left hemisphere and are; Cerebral Cortex, Olfactory areas, Hippocampal region, Striatum, Pallidum, Thalamus, Hypothalamus, Midbrain, Pons, Medulla and Cerebellum (see Supplementary Materials). In a more detailed annotation, each of these regions are divided into sub-regions, comprising a total of 94 brain regions in the left hemisphere we refer to as {\ttfamily{`Fine'}}. For display purposes we show Maximum Intensity Projections of $E(v,g)$---for all genes such projections for coronal, sagittal and axial plane are shown in Fig.~\ref{fig:MIproj}.

For genes in the dated set $G_T$, we create age-selected brain-wide gene expression profile $E_T(v)$ defined by
\begin{equation} 
E_T(v) = \frac{1}{|G_T|}\sum_{g\in G_T} E(v,g),
\end{equation}
where $|G_T|$ is the number of genes in set $G_T$. The over-expression of genes belonging to $G_T$ relative to the average level of expression in $v$ can be captured (in logarithmic scale) by the quantity $L_T(v)$ defined as 
\begin{eqnarray}
\label{eq:LTdefs} 
L_T(v) &=& \log\left(\frac{E_T(v)}{E_{\text{tot}}(v)}\right)\\
E_{\text{tot}}(v) &=& \frac{1}{|G|} \sum_{g=1}^{|G|} E(v,g),
\end{eqnarray}
where $N_G$ is the total number of genes. 

\begin{figure}
\begin{center}
\begin{tabular}{|p{3.5cm}|p{1.0cm}|p{5.5cm}|p{5.5cm}|}
\hline
{\bf{\small{Appearance of clade (number of species)}}}  &  {\bf{\small{Nb of  genes}}} &{\bf{Logarithm of ratio to  {\hbox{average}} expression}} & {\bf{Profile of enriched regions in the ARA {\small{($L_T-{\mathrm{mean}}\geq 2$ stdevs, or best region)}}}}\\ \hline

 {\bf{Eukaryota (38)}}  & 3 &\includegraphics[width=\epsWidth]{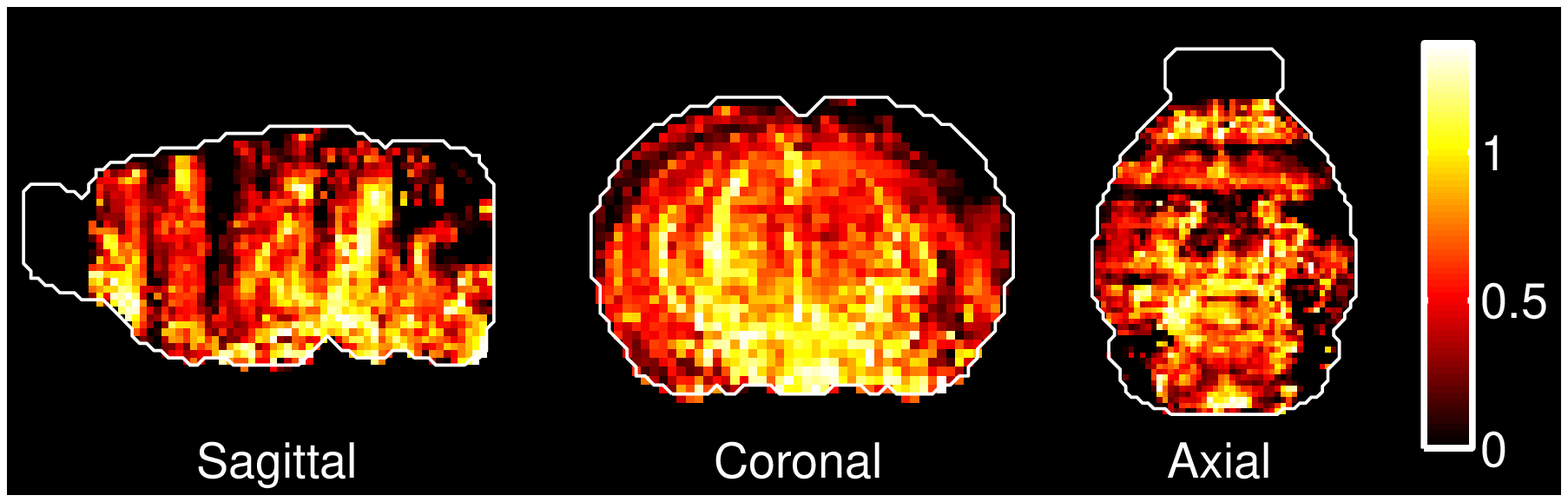}
        & \includegraphics[width=\epsWidth]{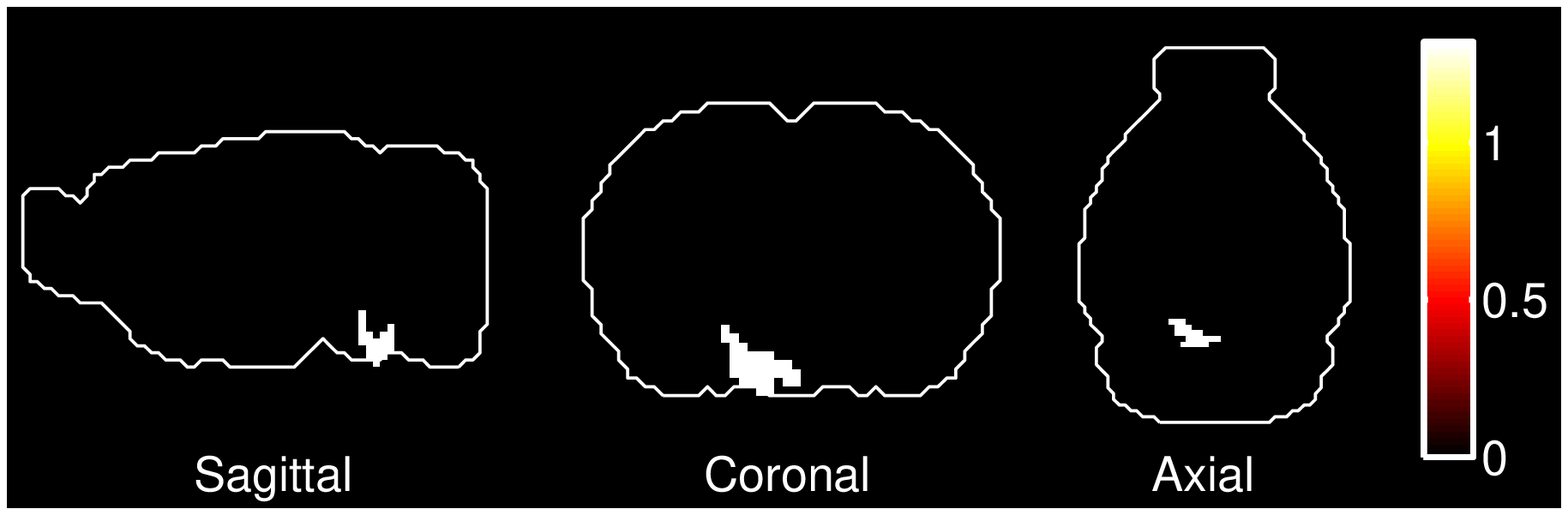} \\ \hline
 {\bf{Metazoa (8)}}  &  159 &\includegraphics[width=\epsWidth]{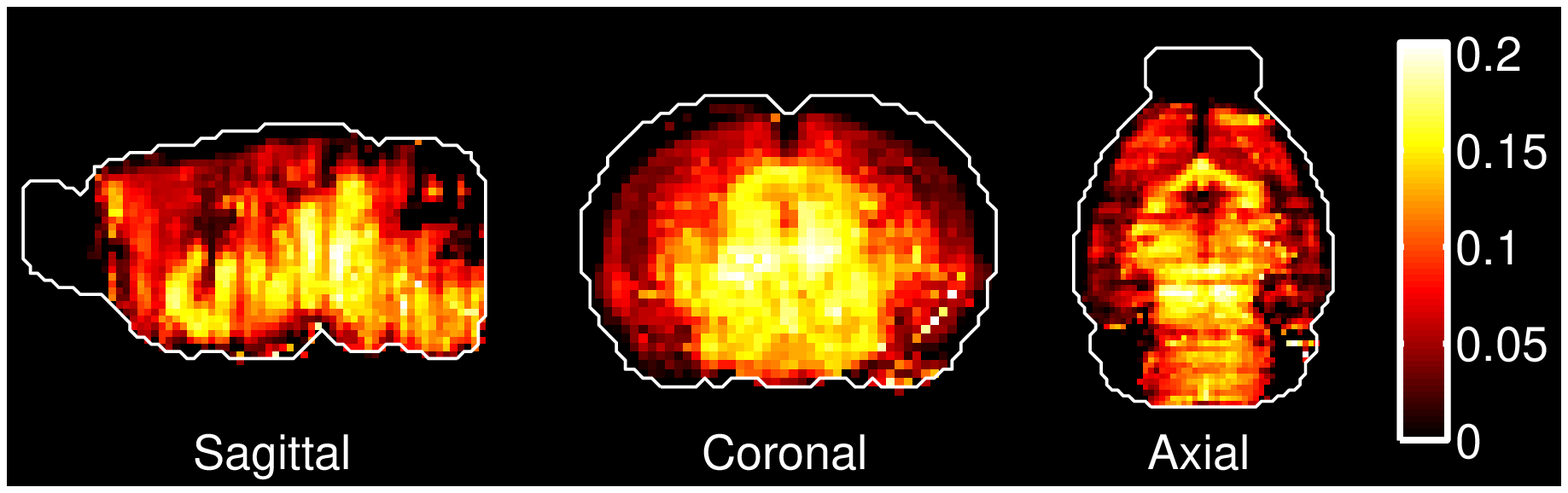} 
         & \includegraphics[width=\epsWidth]{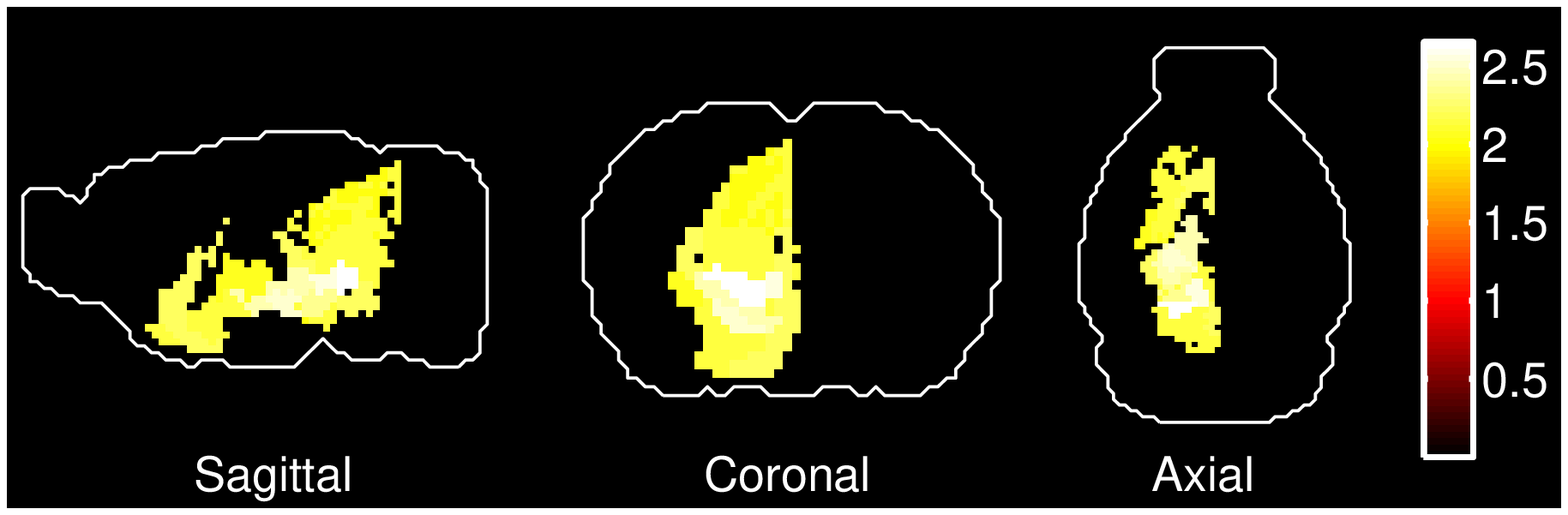} \\ \hline
{\bf{Coelomata  (13)}}  &  52 &\includegraphics[width=\epsWidth]{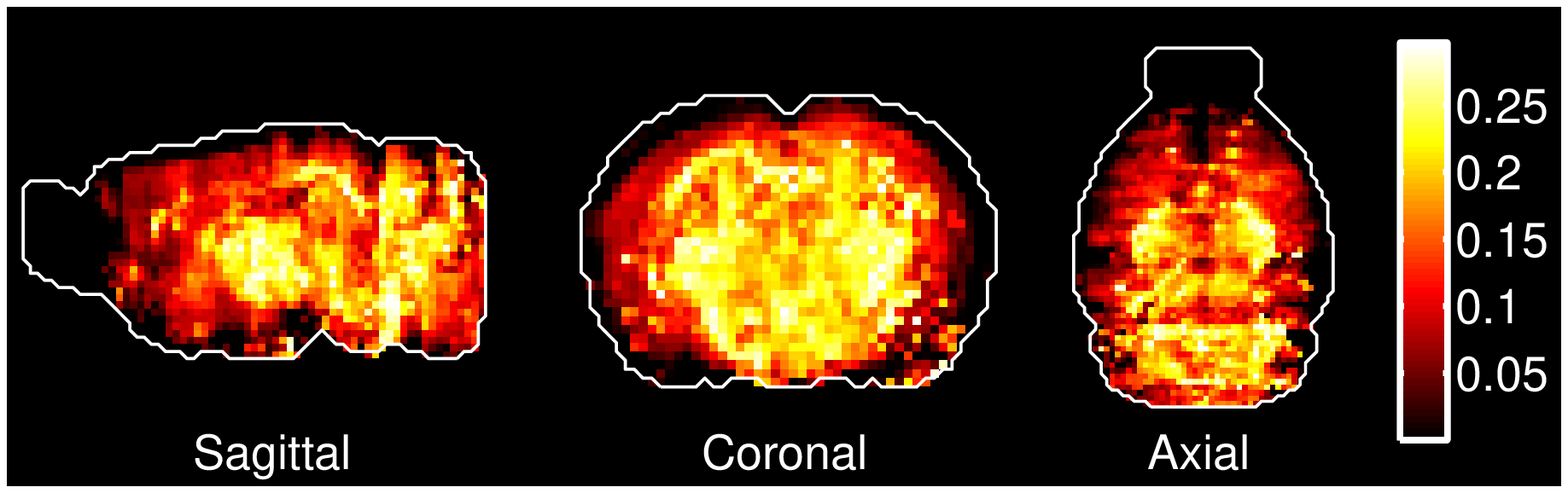}  
        & \includegraphics[width=\epsWidth]{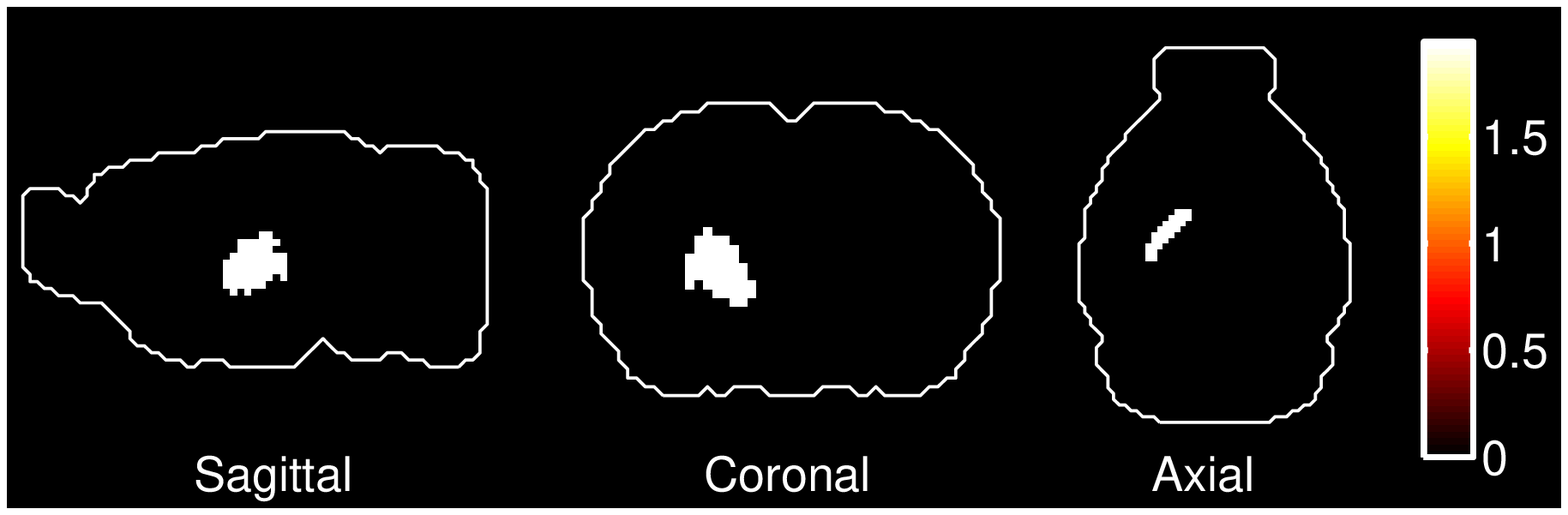}  \\ \hline
{\bf{Chordata  (3)}}  & 88 &\includegraphics[width=\epsWidth]{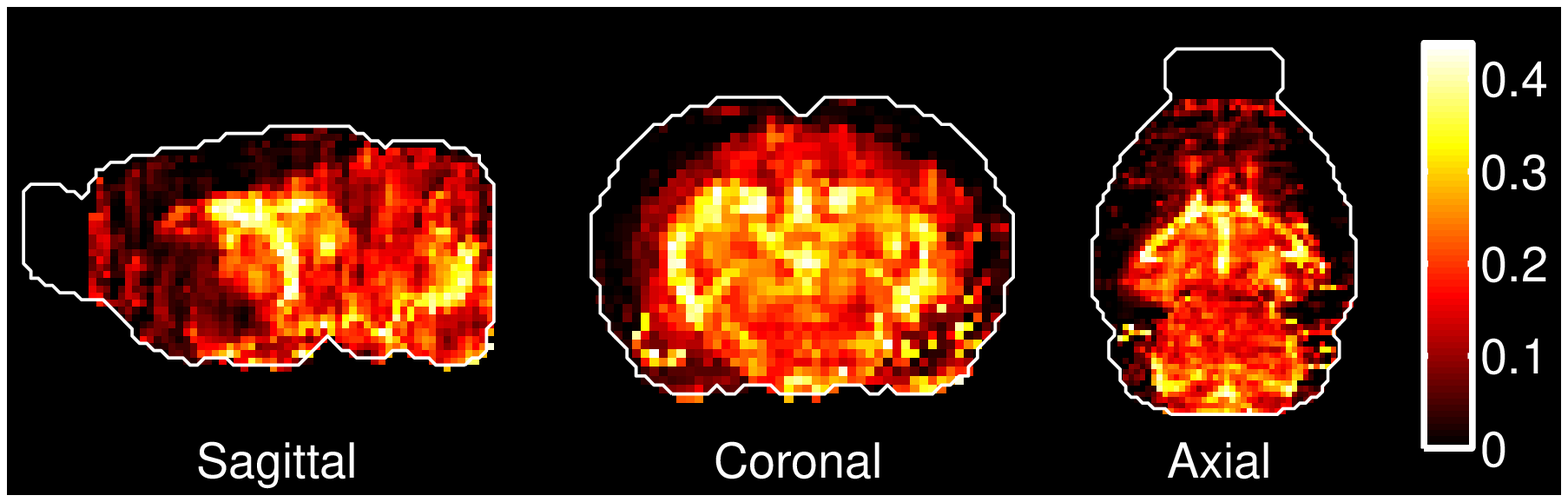} 
           & \includegraphics[width=\epsWidth]{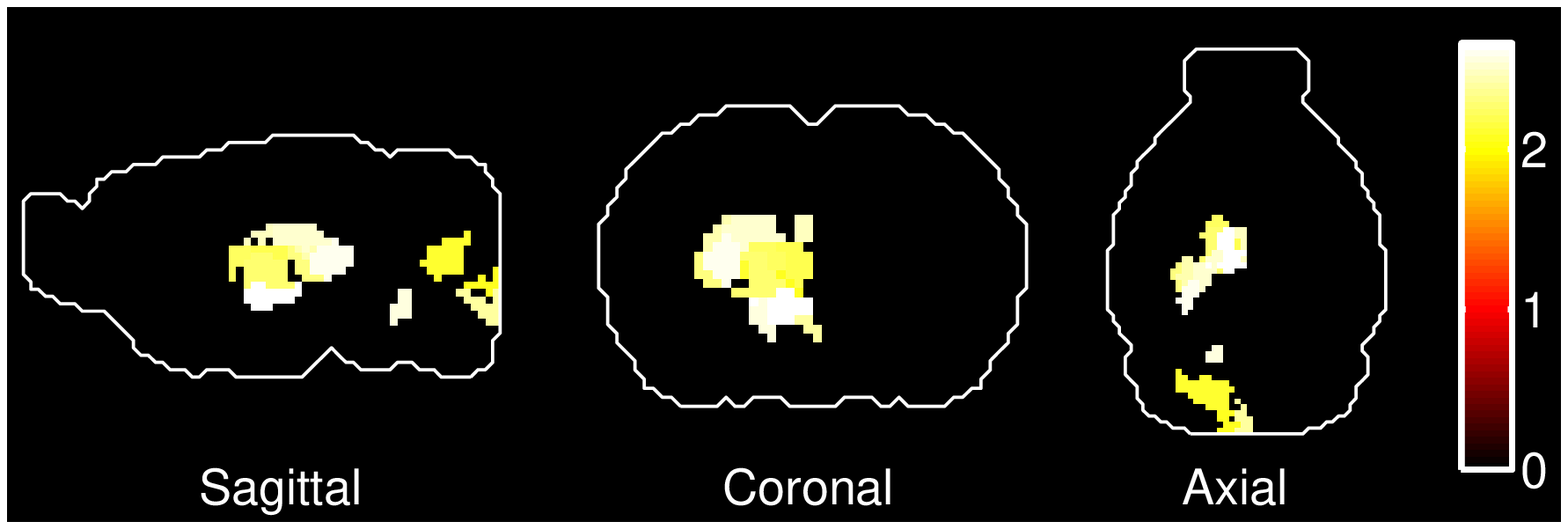} \\ \hline
{\bf{Vertebrata (5)}} & 443 &\includegraphics[width=\epsWidth]{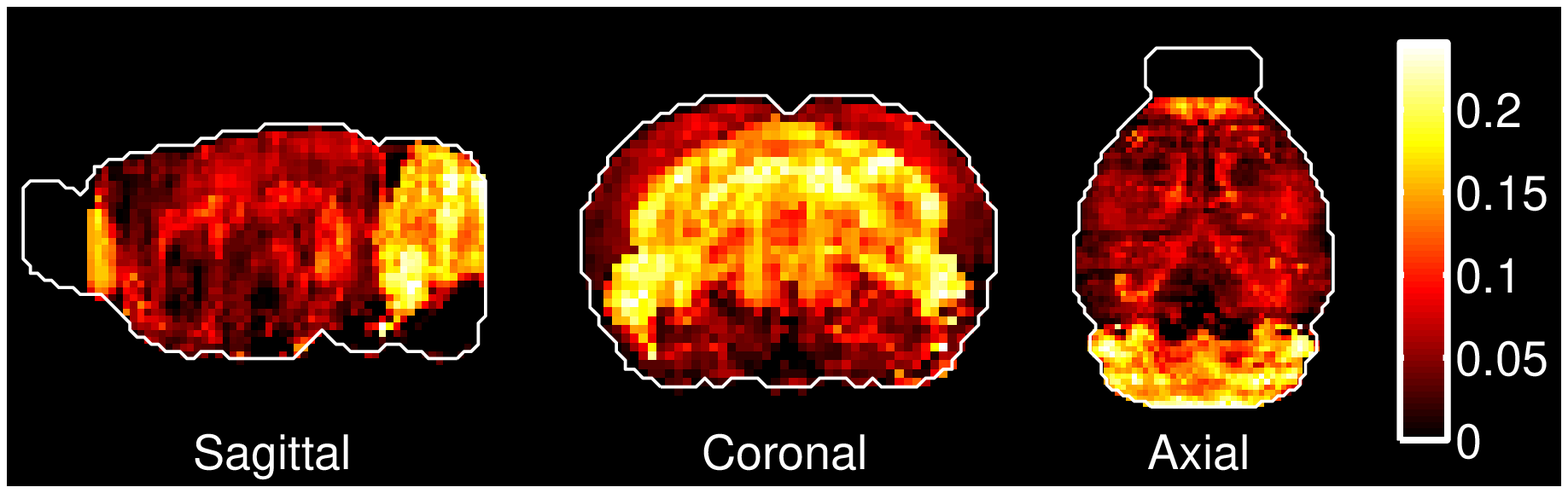} 
          & \includegraphics[width=\epsWidth]{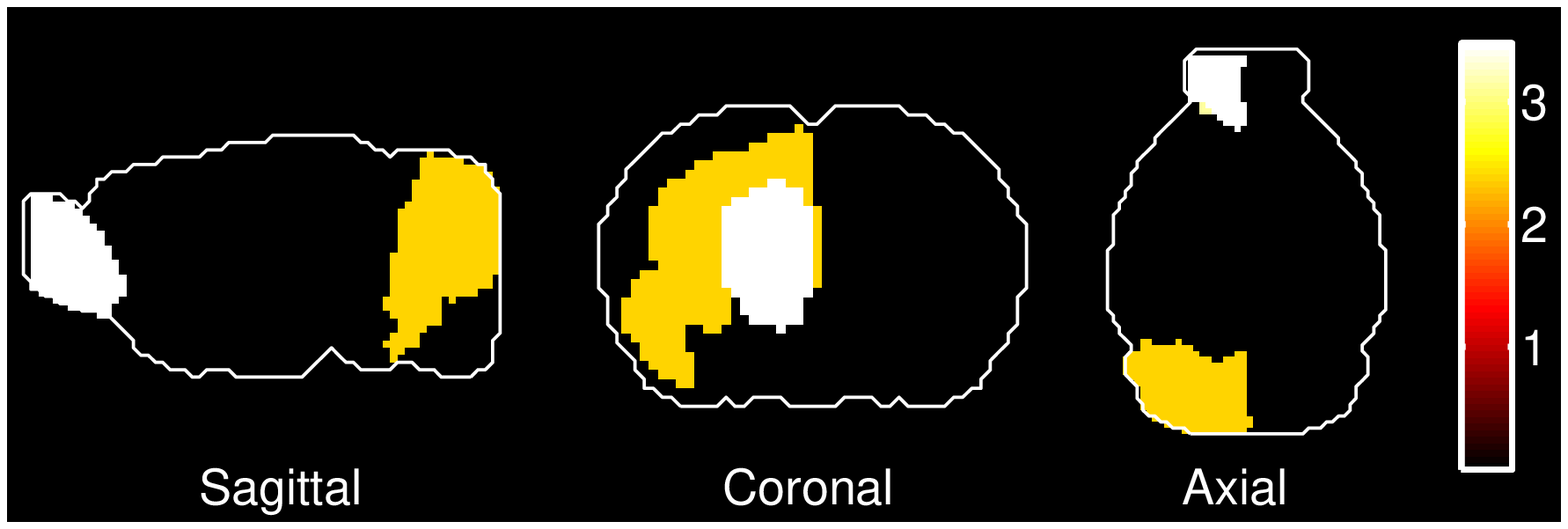} \\ \hline

{\bf{Sarcopterygii (5)}}  & 328 & \includegraphics[width=\epsWidth]{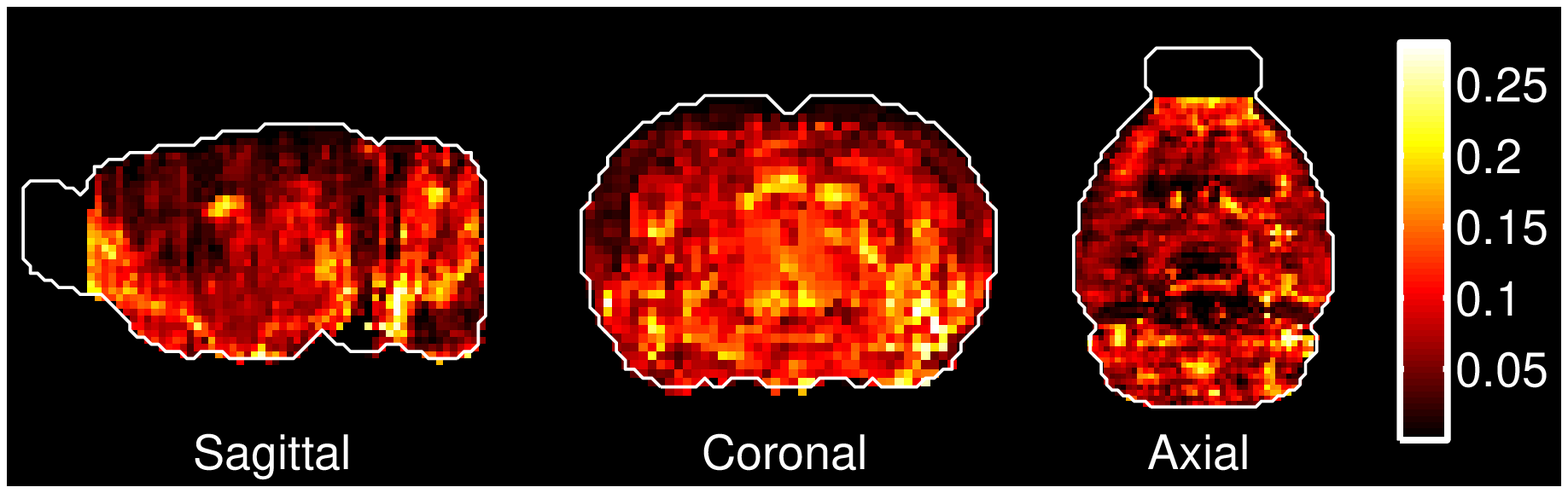} 
         & \includegraphics[width=\epsWidth]{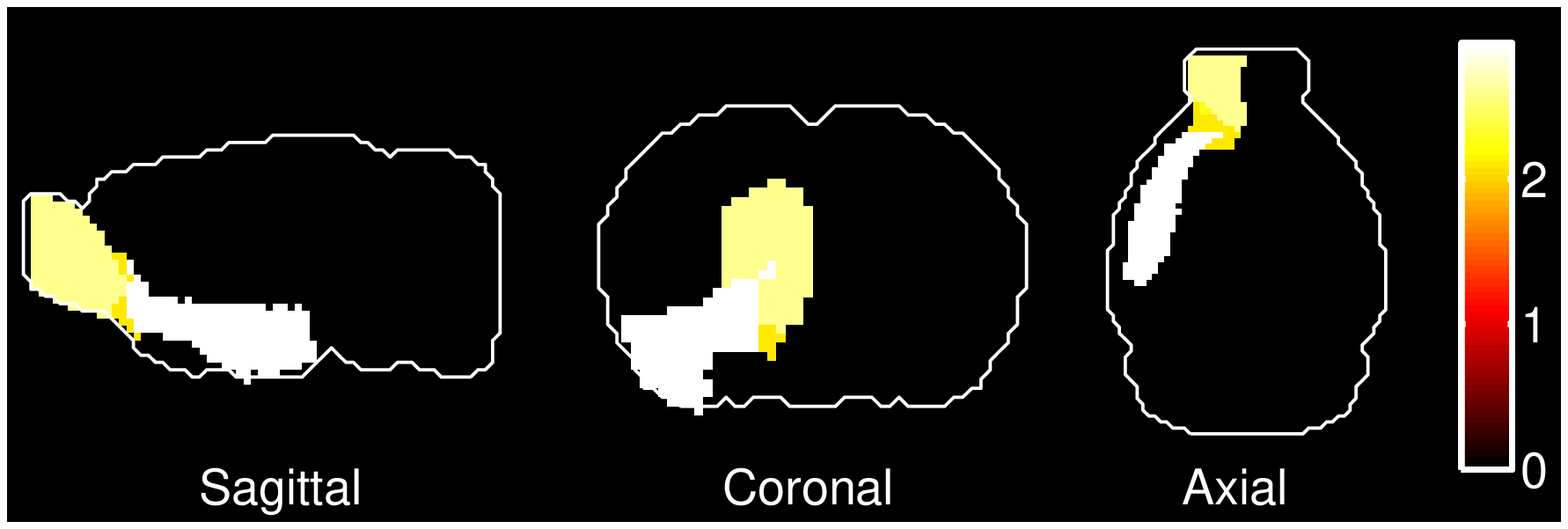} \\ \hline

{\bf{Mammalia (3)}}  & 70 &\includegraphics[width=\epsWidth]{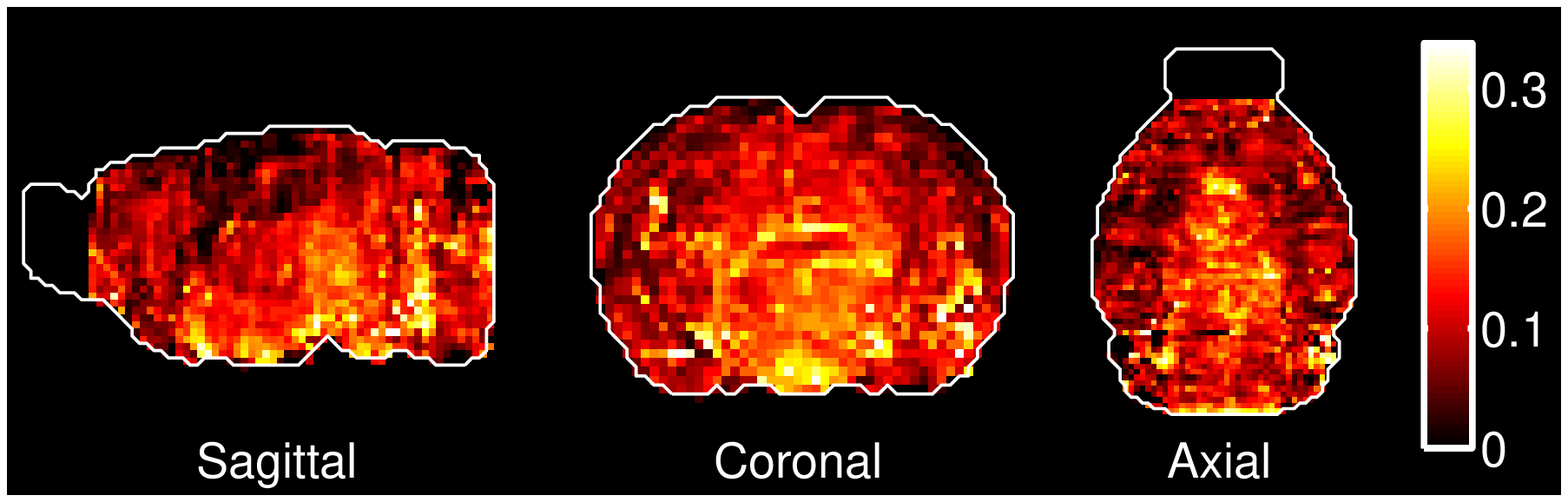} 
              & \includegraphics[width=\epsWidth]{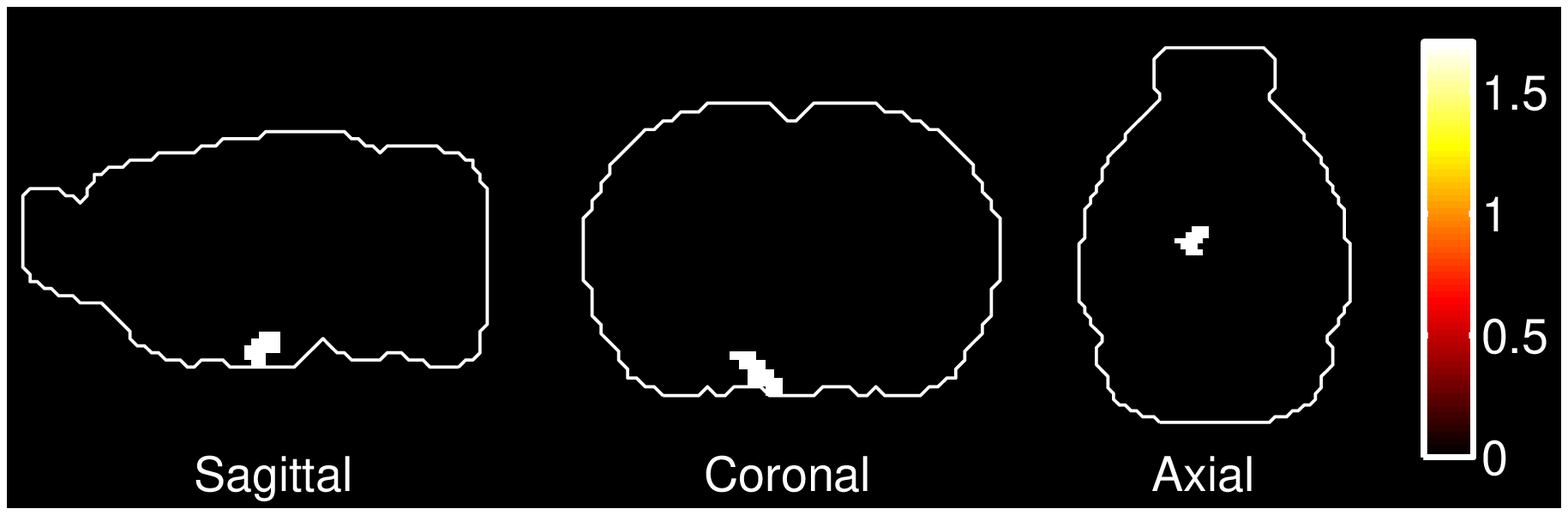} \\ \hline
{\bf{Eutheria  (16)}} &  35 & \includegraphics[width=\epsWidth]{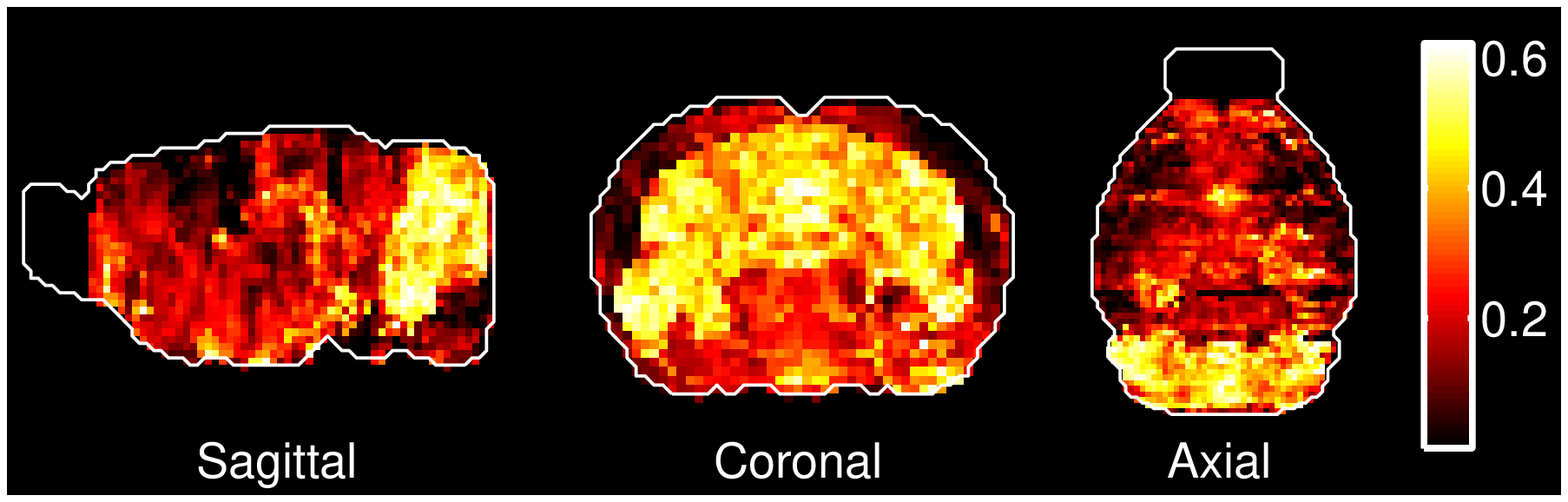}
              & \includegraphics[width=\epsWidth]{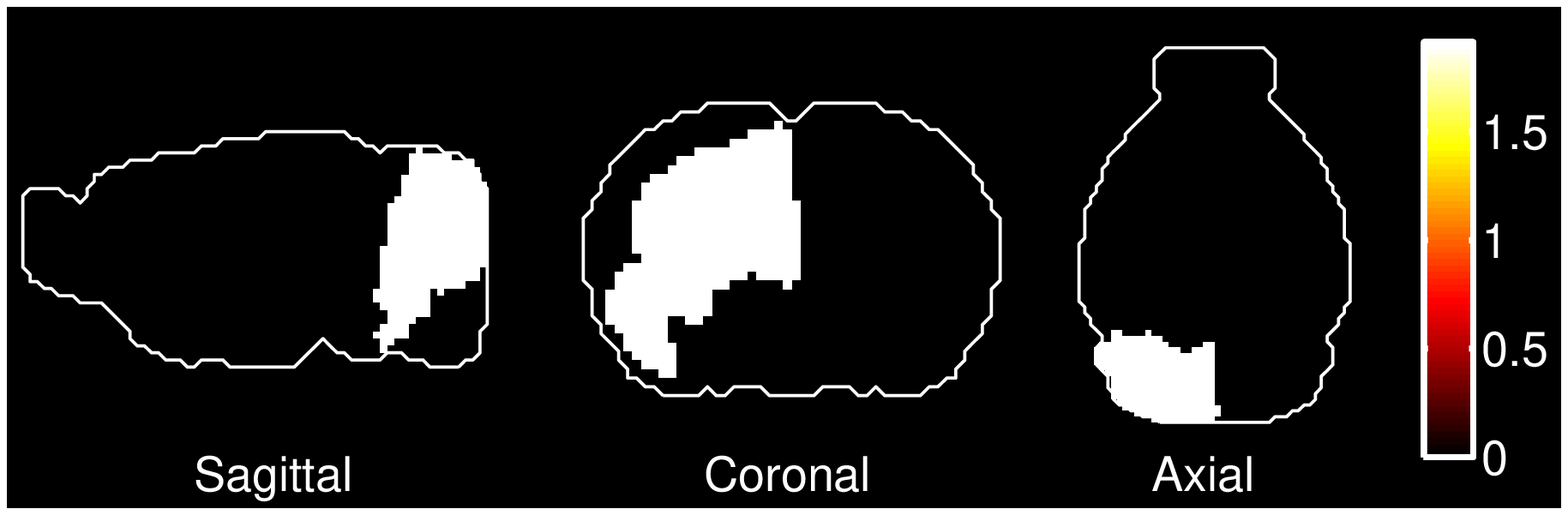}     \\ \hline
{\bf{{\small Euarchontoglires}(12)}} & 47 & \includegraphics[width=\epsWidth]{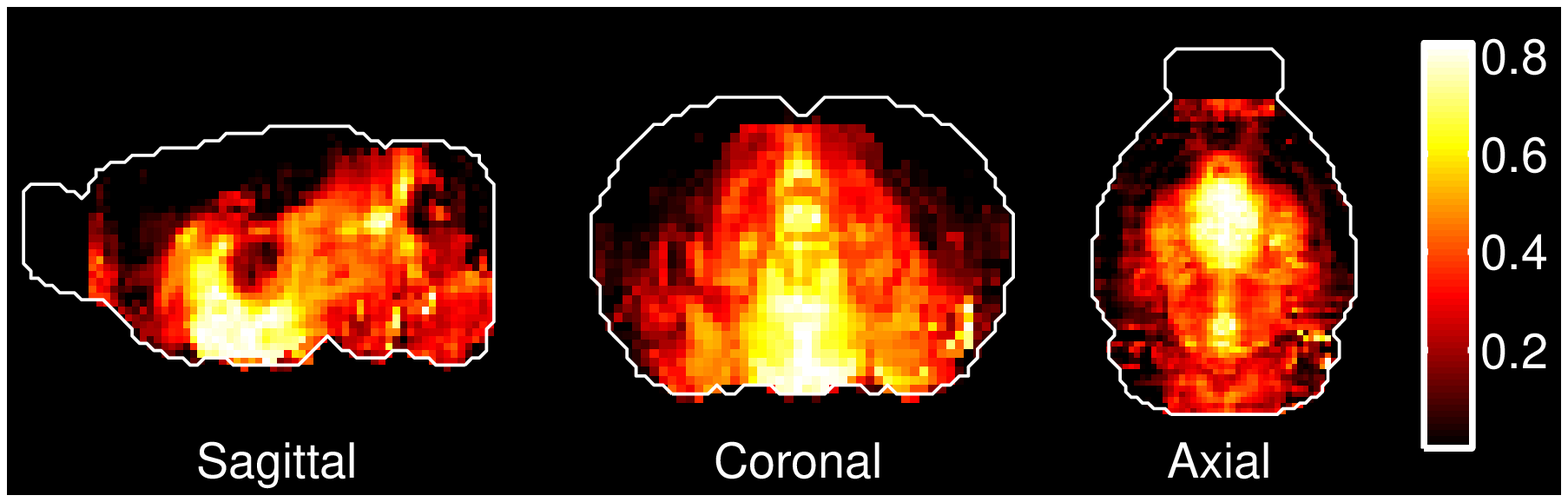} 
              & \includegraphics[width=\epsWidth]{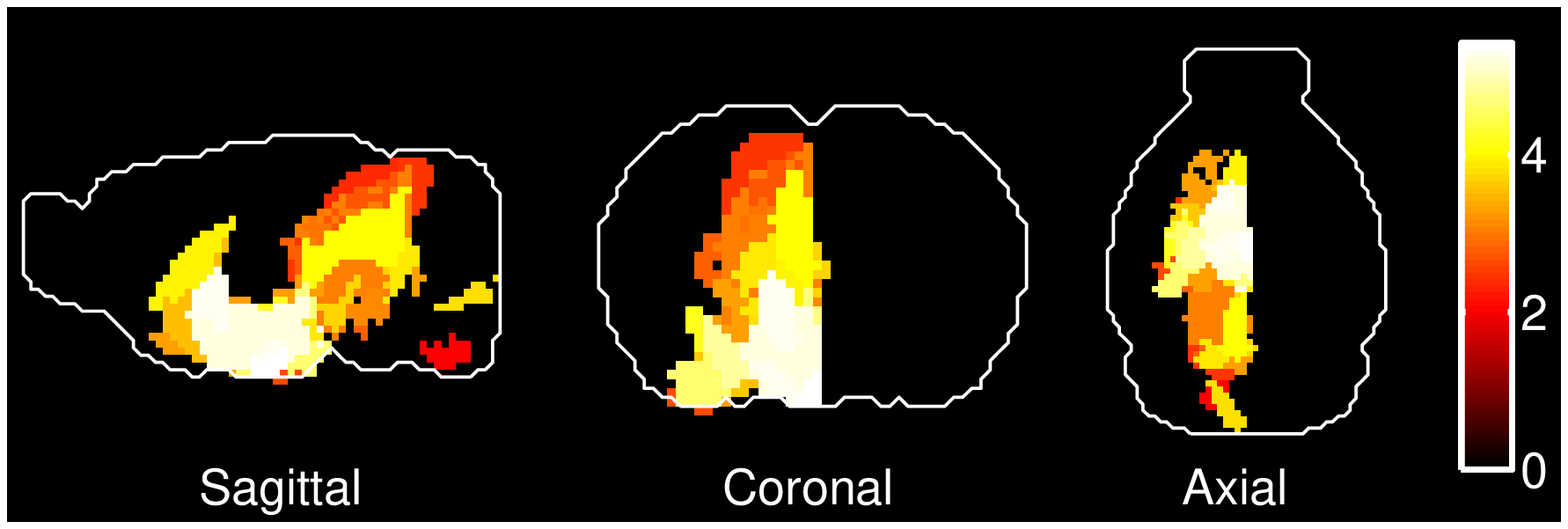}\\ \hline

{\bf{Rodentia (5)}} & 8 &\includegraphics[width=\epsWidth]{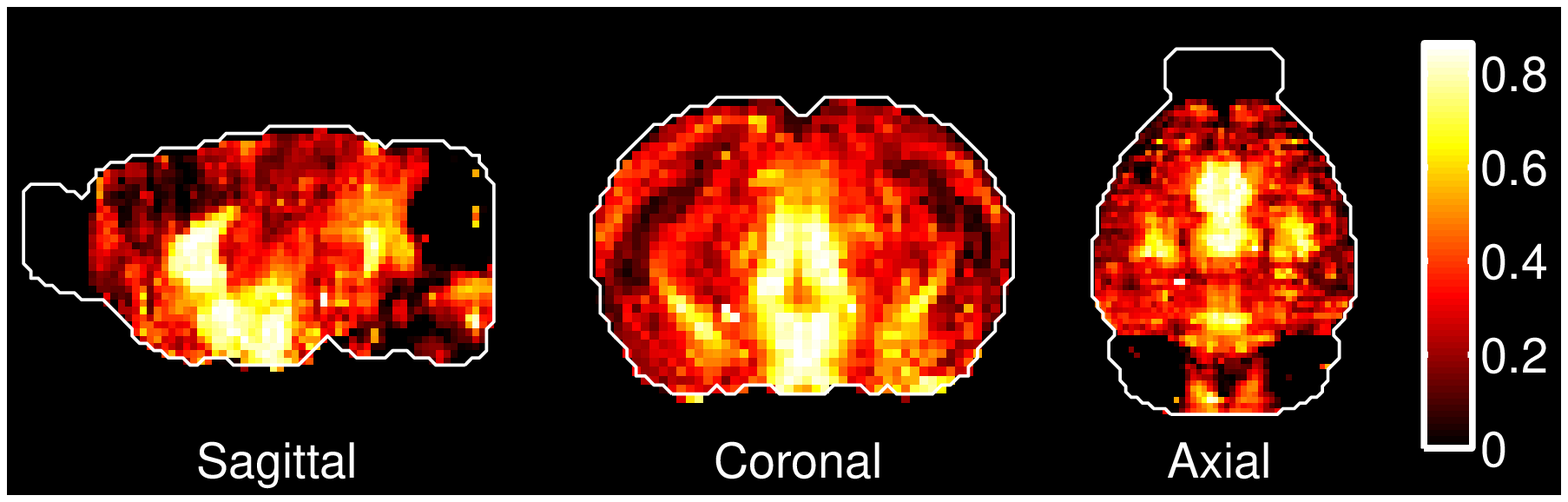} 
             & \includegraphics[width=\epsWidth]{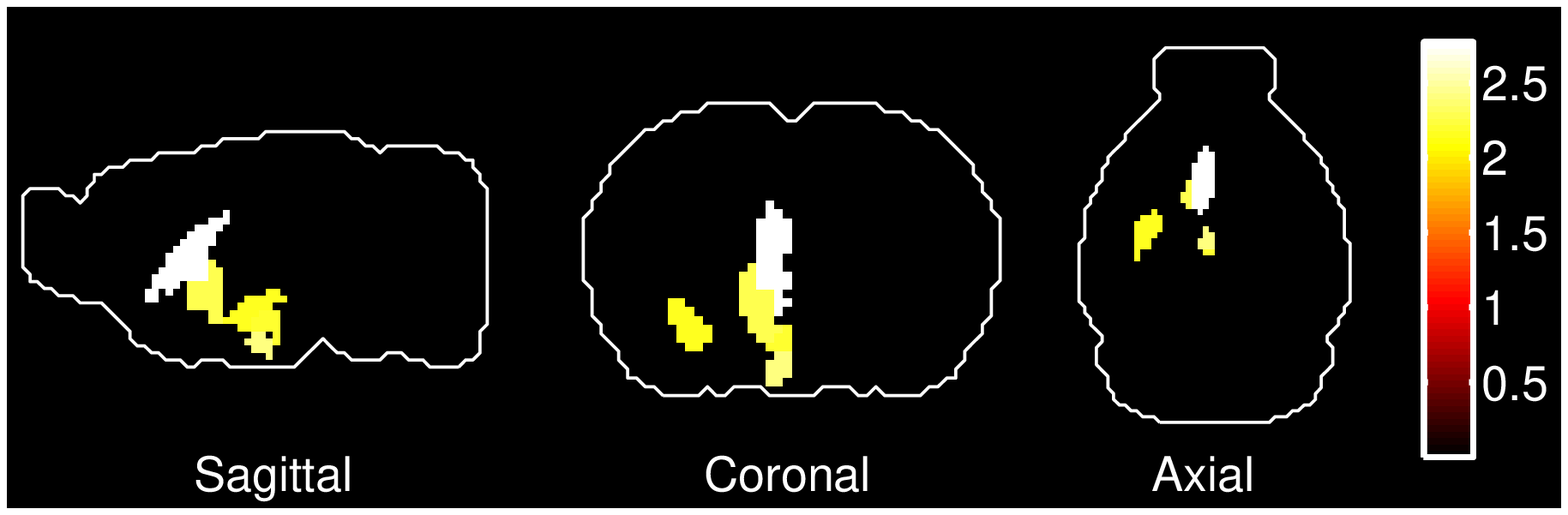}  \\ \hline
{\bf{Mouse (1)}}  & 388 &\includegraphics[width=\epsWidth]{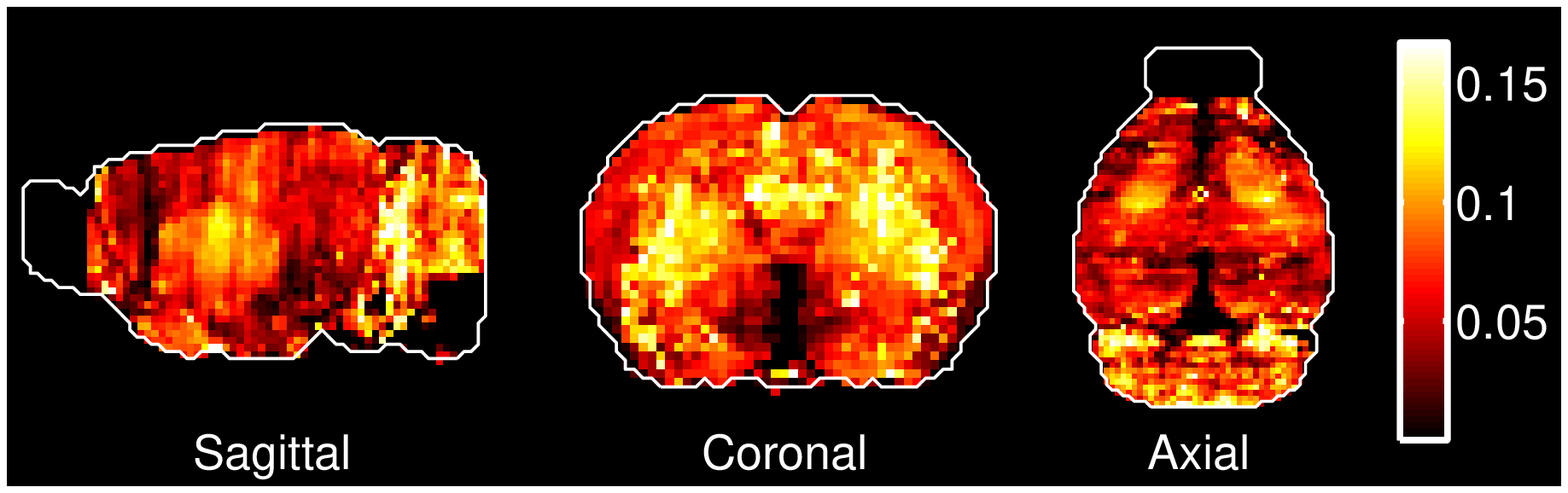} 
            & \includegraphics[width=\epsWidth]{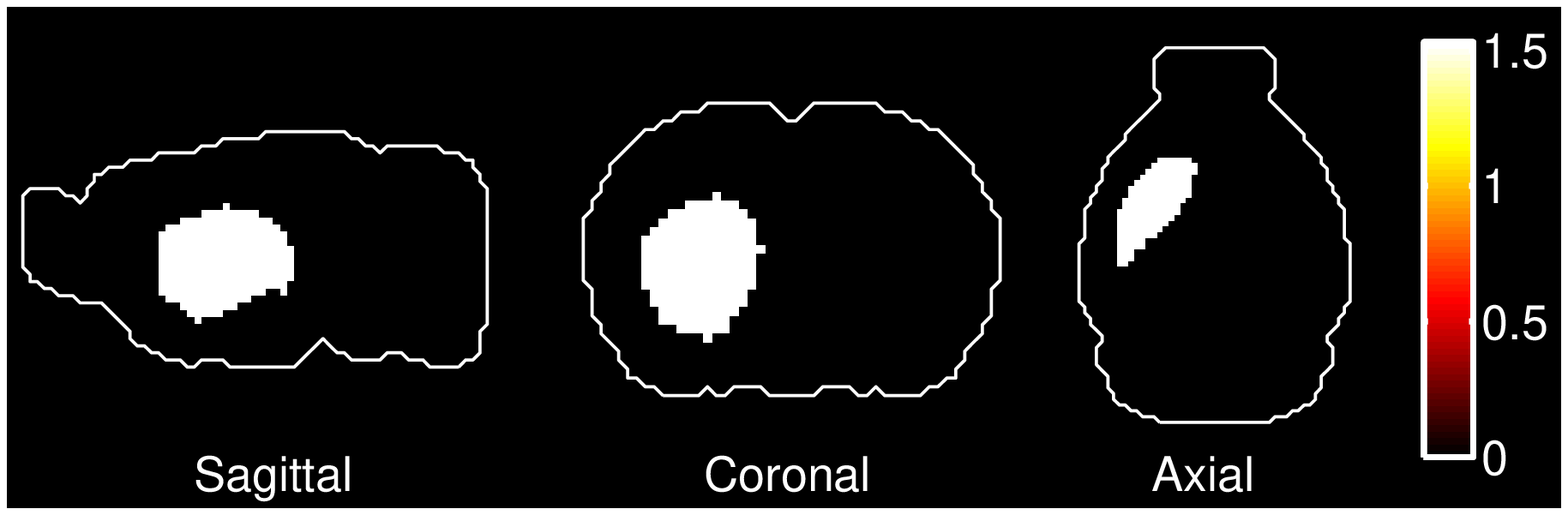} \\ \hline

\end{tabular}
\captionof{figure}{Heat maps of the log-ratios of genes expression of age-selected genes, $L_T$, to the average across the AGEA. Profile of {\ttfamily{`Fine'}} atlas regions (left hemisphere) where $L_T$ is either 2 standard deviations higher than the (bootstrap) mean, or if no such region found, the most significant one is shown.}
\label{fig:logRatio}
\end{center}

\end{figure}

We show the Maximal-Intensity Projections of $L_T$ in Fig.~\ref{fig:logRatio}, summarizing our main results. The patterning of $L_T$ across the brain regions inform us on the signature left by molecular evolution on neuroanatomical regions. Our key observations on the significant over-expression as judged by $L_T$ is summarized in tabular form in Table~\ref{tab:results}, which we discuss in detail in Discussion section. 
\begin{table}[t]
\centering
\begin{tabular}{|l|l|}
\hline
\textbf{Genes novel in clade}& \textbf{Regions significantly over-expressed (significant $L_T$)} \\ \hline 
Eukaryotes & none \\\hline
Metazoa  & Pallidum and Midbrain\\ \hline 
Coelomata & None \\ \hline
Chordata &  Hippocampal region, Thalamus, Pons and Medulla \\ \hline 
Vertebrata & Olfactory areas and Cerebellum \\ \hline 
Sarcopterygii & Olfactory areas \\ \hline
Mammalia & Hypothalamus \\ \hline 
Eutheria & Cerebellum \\ \hline 
Euarchontoglires & Pallidum, Hypothalamus, Midbrain, Pons, Medulla \\ \hline 
Rodentia & Hypothalamus, Striatum\\ \hline
\end{tabular} 
\caption{Regions where dated genes grouped by clade is over-expressed with statistical significance (two standard deviation or more).} 
\label{tab:results} 
\end{table} 

What are the attributes of the genes that contribute significantly to the over-expression in a specific brain region at a specific timescale? The list of dated genes significantly over-expressed in a specific brain region can be large, and we need to rank their importance. This is done by computing the localization score $\lambda(g, R)$ of genes $g$ in brain region $R$, see Ref.~\cite{Grange2012} for further details. The definition of $\lambda(g,R)$ is as follows---
\begin{equation}
\lambda(g, R) = \frac{\sum_{v\in R} E(v,g)^2}{\sum_{v \in \Omega} E(v,g)^2}, 
\end{equation}
where $\Omega$ denotes the whole brain. The score $\lambda(g,R)$ is a positive number lesser than one---genes with expression contained within a specific brain region enjoy a high localization score. For the regions in Table~\ref{tab:results} corresponding to the clades, we report the attributes of the dated genes with high-localization scores $\lambda(g,R)$ in those regions. These attributes are obtained from Mouse Genome Informatics database~\cite{MGI}. The functional annotation/names of genes is reported in Table~\ref{tab:results2}, and is primarily meant as a practical summary of an otherwise long list of diverse genes. A comprehensive list can be found in Supplementary Material and see further discussion of Table~\ref{tab:results} and ~\ref{tab:results2} in Discussion section.  

\begin{table}[t]
\centering
\begin{tabular}{|l|l|}
\hline
\textbf{Genes novel in clade}& \textbf{Gene attributes of high localization-score genes in regions of Table~\ref{tab:results}} \\ \hline 
Eukaryotes & none \\ \hline 
Metazoa  & chloride channel, GABA transporter, chemokines, sodium/calcium exchange, \\ 
{} &        potassium channel \dots            \\ \hline 
Coelomata & None \\ \hline
Chordata &  Retenoic acid receptor, contactins, growth differentiation factor, calcium channel,  \\
{} &      potassium channel, otic morphogenesis, glial fibrillary acidic protein, anion exchanger, \\
{} &      peripherin, actin cytoskeletal related proteins, frizzled-related proteins \dots \\ \hline 
Vertebrata & Synaptotagmin, versican, calcium channel voltage-dependent subunits, \\
{} & frizzled-related proteins, corticotropin releasing hormone, \\
{} & receptor (calcitonin) activity modifier, cadherin-related, semaphorins, \\
{} &  Adenomatous polyposis coli (APC), olfactory receptors, glutamate receptor, \\
{} &   ephrin receptor, GABA receptor subunit,  neurogenic differentiation, opsins,  \\
{} & otic morphogenesis, edothelin receptors, canabinoid receptor, Cholinergic receptor,  \\  
{} & visual system homeobox \dots \\ \hline 
Sarcopterygii & RAS activator, neuron differentiation, thyrotropin releasing hormone receptor,\\
{} &  semaphorins, complexin, potassium voltage-gated channel \dots \\ \hline 
Mammalia & pregnancy-upregulated kinase, cytokine receptor, neurexophilin \dots \\ \hline 
Eutheria & Purkinje cell protein, cardiotrophin, titin \dots \\ \hline 
Euarchontoglires & insulin receptor substrate, GABA receptor subunit, \\
{} &  thyrotropin releasing hormone, huntingtin-associated protein, neuronatin, necdin, \\
{} & nerve-growth factor receptor, calcitonin-related, growth-hormone-releasing,\\
{} & chloride channel \dots \\ \hline
Rodentia & metal ion transporter, trophinin \dots \\ \hline
\end{tabular} 
\caption{Summary of annotation of genes with high-localization scores in regions listed in Table~\ref{tab:results}.} 
\label{tab:results2} 
\end{table}

\section*{Materials and Methods} 
\subsection*{Determining the age of genes} 

Algorithms and biologically relevant criteria for determining gene orthologs through sequence alignment have been discussed extensively in the literature~\cite{Yang1997, Wall2003, Wall2007, Kuzniar2008,Yang2007, Zhou2007, Altenhoff2009, Chen2007}. Gene fusion/fission events leads to non-orthologous genes sharing similar protein domains, therefore, the query sequence is usually mapped to the longest transcript (protein sequence) and a minimum-length-criterion on the sequence alignment is imposed for orthology. A tool like protein-BLAST is usually used to search for similar sequences across the whole genome. However, sequence similarity is not sufficient in establishing orthology and various algorithms and databases have been developed for reliable genome-wide orthology~\cite{Chen2006a, Chen2007, Deluca2006, Datta2009, Vilella2009, Ostlund2010}. For example, reciprocal gene loss events in two species may lead to the lone surviving paralogs in each species to be detected as orthologs, and therefore (an amino-acid substitution rate based) distance criteria is necessary to filter sequence similarity results~\cite{Altenhoff2009, Wall2007, Wall2003}. Though earlier work on neuronal genes~\cite{Ryan2009} have depended solely on BLAST tools, we use the OMA dataset~\cite{Schneider2007, Roth2008, Altenhoff2011} comprising a comprehensive list of all orthologs found for all fully-sequenced species. Our main reasons for choosing this dataset are
\begin{itemize}
\item{The computational rigor employed in determining sequence similarity and declaring orthology. The sequence alignment tool is Smith-Waterman, a relaxed Reciprocal Smallest Distance criterion~\cite{Roth2008, Wall2007} using the evolutionary distance between sequences is employed to determining candidate orthologs, and triangle equalities for pair-wise distances between such candidates of the two query species as well as a third `witness' species (that has older common ancestor) is used to filter potential orthologs.} 
\item{An up-to-date analysis of all pairwise orthologs of all fully-sequenced eukaryotes is available.}
\end{itemize}

\begin{figure}[ht]
	\begin{center}
		\includegraphics[width=6in]{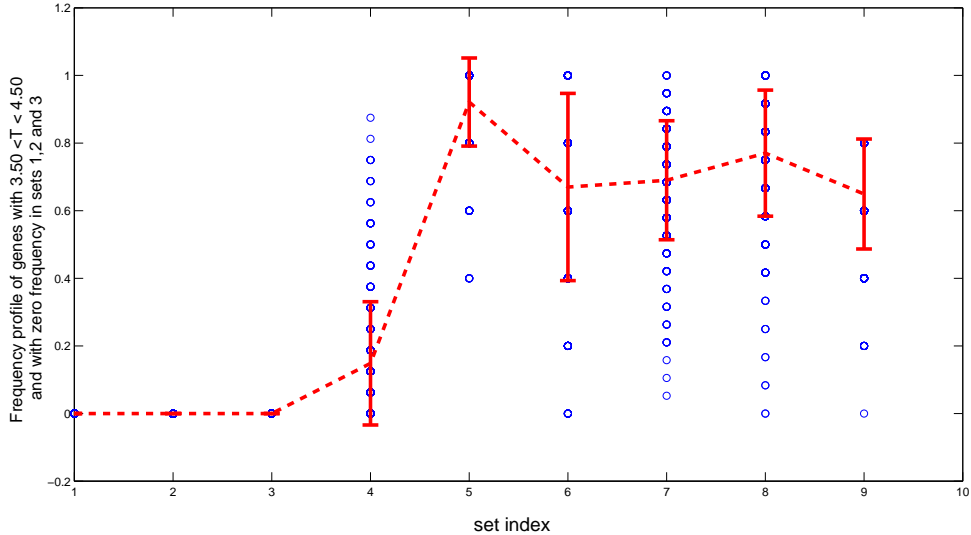}
	\caption{Average profile of genes novel in fourth clade}
	\label{fig:profile4} 
	\end{center}
	
\end{figure}

To determine the `age' of a gene we proceed as follows. We use the order of clades introduced in the Results section. Denote this ordering by $S_T$;  the set of species that belong in clade $T$. Consider the species in the set difference, $\Delta_{T} = S_T - S_{T+1}$. Orthologs of a gene $g$ in this set $\Delta_T$ bear witness to whether $g$ was present in the common ancestor of $S_T$. Naively, one expects to observe a sudden appearance of $g$ in $\Delta_T$ for some $T$, and its ubiquitous presence subsequently. However, gene loss events may lead to the disappearance of a gene from an entire branch of the phylogenetic tree. Moreover, because we can only sample a fraction of all speciation events in any evolutionary history, gene orthologs may appear in only a \emph{few} of all the species in $\Delta_T$; the ones that share a common ancestor with $S_{T+1}$. Thus, the frequency of finding orthologs in $S_T$ show signature of both invasion of a gene and its loss on branches of the tree. We fit the frequency profile to a phenomenological function with three parameters,  
\begin{equation}
f(s) = \frac{\exp\left[-\mu\,(s - T)\right]}{ 1 + \exp\left[-(s-T)/b\right] },
\end{equation}
where $T$ is the `age', $\mu$ corresponds to the loss rate, $b$ the rate of invasion, and $s$ indexes the ordering of clades.  Further discussion are relegated to Supplementary Material. As an example, the average profile of a few random genes novel in the fourth clade is shown is plotted in Fig.~\ref{fig:profile4}. 

For our analysis, we round the fitting parameter $T$ to obtain the discrete `age' score for each gene. 


\subsection*{Analysis of gene expression} 

In our results section we report $L_T(v)$ of genes grouped by `age' $T$. Here we present analysis of the statistical significance of the patterns observed in $L_T(v)$ using a bootstrap strategy. 

\begin{figure}[ht]
	\begin{center}
		\includegraphics[width=5in,keepaspectratio]{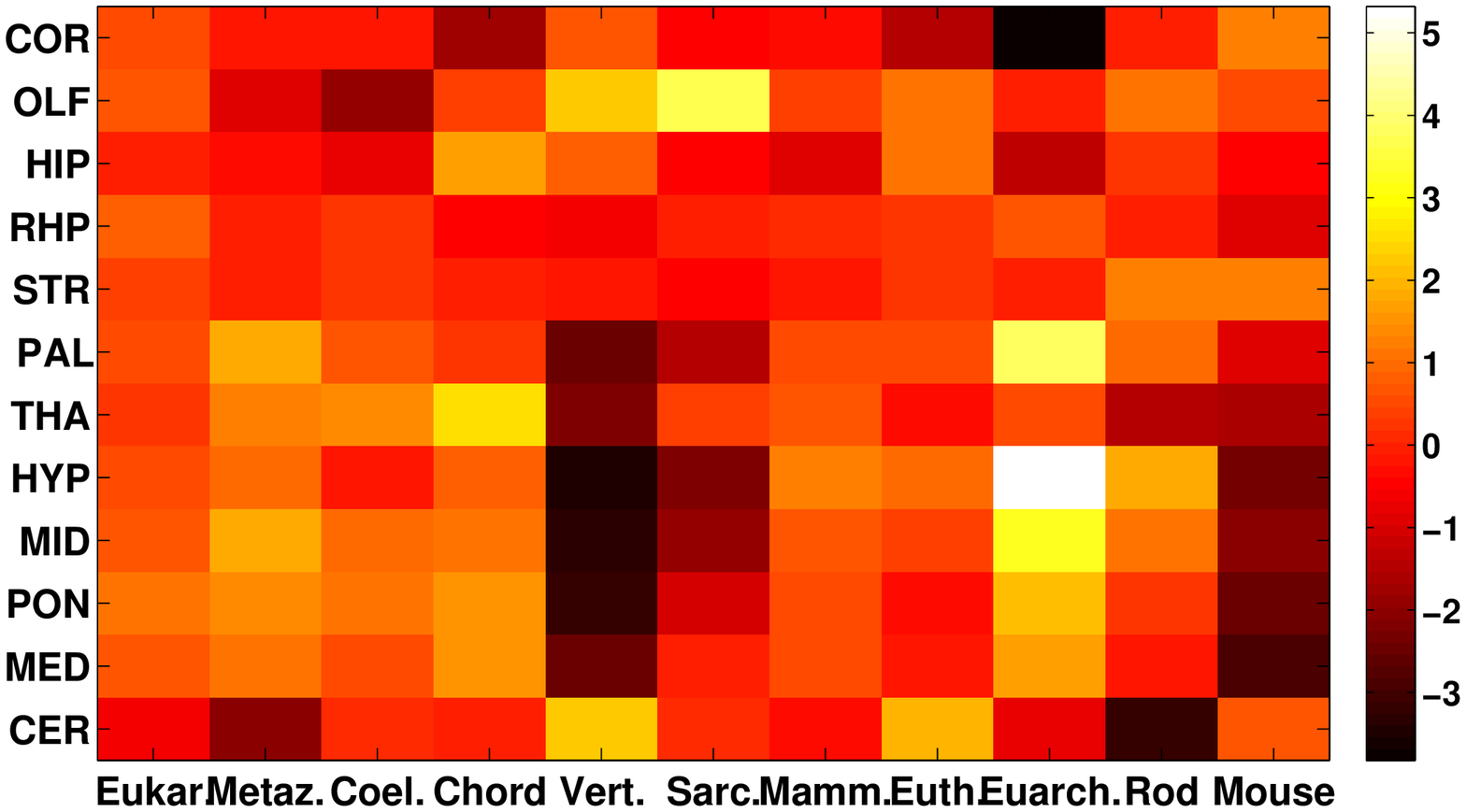}
	\caption{Heatmap summarizing significance of observed pattern in $L_T$, determined by bootstrap analysis. The abbreviation in the y-axis:  COR = Cerebral cortex, OLF = Olfactory areas, HIP = Hippocampal region, RHP = Retrohippocampal region, STR = Striatum, PAL = Pallidum, HYP = Hypothalamus, THA = Thalamus, MID = Midbrain, PON = Pons, MED = Medulla, CER = Cerebellum. The abbreviations is the x-axis are for the clade ordering: EUK = Eukaryota, MET = Metazoa, COE = Coelomata, CHO = Chordata, VER = Vertebrata, SAR = Sarcopterygii, MAM = Mammalia, EUT = Eutheria, EUA = Euarchontoglires, ROD = Rodentia, MOU = Mouse.}
	\label{fig:HeatMapsStd} 
	\end{center}
	
\end{figure}

Denote the size of the set $G_T$ by $|G_T|$. We consider the log-relative gene expression energy $L_T(R)$ for each of the {\ttfamily{`Big12'}} brain parcellations by $R \in [1,12]$ where the regions are \{Cerebral Cortex, Olfactory areas, Hippocampal region, Retrohippocampal region, Stratum, Pallidum, Thalamus, Hypothalamus, Midbrain, Pons, Meduall, Cerebellum \}.  We repeatedly draw (ten thousand draws) a random set $S_{\text{rand}}$ of $|G_T|$ number of genes from the total number of genes $G$ (=3041) and compute $L_T(R)$ given by equations equivalent to Eqs.~\ref{eq:LTdefs} ---
\begin{eqnarray}
L_T(R) &=& \log\left( \frac{E_T(R)}{E_{\text{tot}}} \right)\\
E_T(R) &=& \frac{1}{|G_T|} \sum_{g\in S_{\text{rand}}} \sum_{v \in R} E(v,g) \\
E_{\text{tot}} &=& \frac{1}{|G|} \sum_{g=1}^{|G|}  \sum_{v \in R} E(v,g)
\end{eqnarray}
Comparison of the fluctuations in regions specific log-relative gene expression energies $L_T^{\text{rand}}(R)$ computed for random draws against the fluctuation in $L_T(R)$ computed for the `age'-specific genes $G_T$ for each $T$, informs us on the signal to noise ratio and the significance of the patterns we observe. A sample plot of $L_T(R)$ and $L_T^{\text{rand}}(R)$ against $R$ for the genes novel in the Euarchontoglires is shown Fig.~\ref{fig:NonRodentia_bootstrap} (also see Supplementary Material). For this example, the pattern observed in $R=9,10,11$ i.e., Midbrain, Pons and Medulla are at least three standard deviation stronger than expected by chance. Moreover, some of he striking neuroanatomical patterns on the heat maps in Fig.~\ref{fig:logRatio} are confirmed at the level of two standard deviations, for instance the high expression across cerebellum in genes novel to Eutheria, and the high expression in thalamus, hypothalamus and midbrain in genes novel to Euarchontologlires.
\begin{figure}[ht]
	\begin{center}
		\includegraphics[width=5in,keepaspectratio]{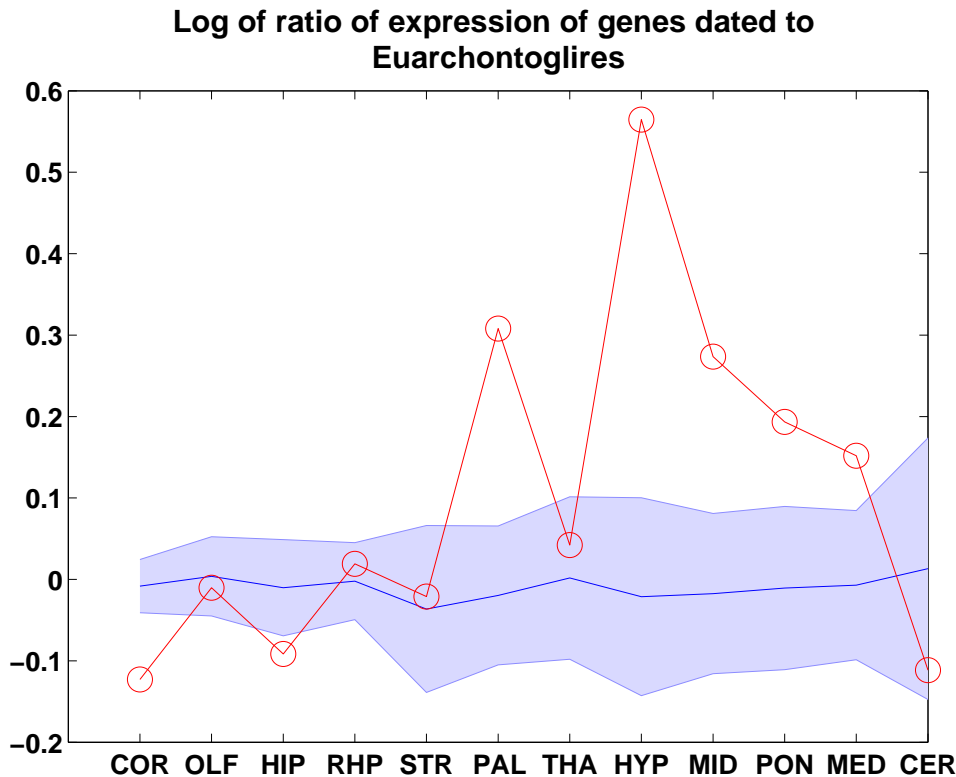}
	\caption{Example of bootstrap analysis to estimate significance of observed log-ratio gene expression pattern of dated genes. The blue line and the blue band are the mean and one standard deviation off mean respectively of $L_T^{\text{rand}}$, the expression profile of randomly selected identical number of genes as $|G_T|$. The red line is observed $L_T$ for $T$ corresponding to Euarchontoglires. The abbreviation in the x-axis: BasicCOR = Cerebral cortex, OLF = Olfactory areas, HIP = Hippocampal region, RHP = Retrohippocampal region, STR = Striatum, PAL = Pallidum, HYP = Hypothalamus, THA = Thalamus, MID = Midbrain, PON = Pons, MED = Medulla, CER = Cerebellum.}
	\label{fig:NonRodentia_bootstrap} 
	\end{center}
	
\end{figure}
We note that the deviations from the mean can be larger in units of standard deviations when we performed the same analysis using the {\ttfamily{`Fine'}} atlas of the left hemisphere that consists of 94 (non-hierarchical) regions. The analogous maps of $L_T$ as a function of the fine-anatomy labels loses clarity in display, therefore we report the results as a heat map of statistical significance. The step taken in the process are as follows:
\begin{itemize} 
\item{Compute $L_T(R)$  and $L_T^{\text{rand}}(R)$ for all $R$ and all $T$.}
\item{For each  `age' $T$ and region $R$, compute 
\begin{equation}
\delta(T,R) = \frac{L_T(R) - {\text{mean}} ( L_T^{\text{rand}}(R) ) }{\text{std}( L_T^{\text{rand}}(R) )}
\end{equation}
where the mean and standard deviation are over the samples created by bootstrapping. }
\item{For each age $T$, determine region $R$ having values of $\delta(T,R)> \delta_\text{crit}$, where we display results for $\delta^\text{crit} = 2$. These regions are the $\delta^{\text{crit}}$-best regions.}
\item{For each age $T$, we have a set $\delta^{\text{crit}}$-best regions. Construct heat map $H_T(\delta^{\text{crit}})$ according to the rule that if a voxel belongs to one of the $\delta^{\mathrm{crit}}$-best regions color the voxel proportional to $\delta(T,R)$, otherwise color it black. For $T$ where no $\delta^{\mathrm{crit}}$-best region is found report the region $R$ with the highest $\delta(T,R)$. }
\end{itemize} 
Maximal-intensity projections of these heat maps $H_T(\delta^{\text{crit}})$ employing the above rules were reported in Fig.~\ref{fig:logRatio}. 
In Fig.~\ref{fig:HeatMapsStd} we summarize the results of bootstrap analysis for the {\ttfamily{`Big12'}} regions where a heatmap of $\delta(T,R)$ is shown for each clade and for each region. The heatmap quantify the significance one can attach to observed patterns in Fig.~\ref{fig:logRatio}. 

\section*{Discussion}

We discuss the implication of the observations summarized in Table~\ref{tab:results} and ~\ref{tab:results2}. The current work serves as a resource determining classification of AGEA genes by evolutionary ages, determining statistical significance of over-expression in brain regions and rank of localization in those regions. The resource can be used for specific inquiry on genes of interest from a functional standpoint. Instead, we discuss broader implications of our findings.   

The cerebellum is over-represented for genes dated to Vertebrata. This is consistent with the fact that hagfishes, which are rather elementary vertebrates (lacking a vertebral column but possessing a skull) lack cerebellum. Moreover, jawless fishes like lampreys, which are ancient vertebrates, have very small cerebellum compared to cartilaginous fishes, which are more recent vertebrates~\cite{ButlerBook2005}. With the exception of mormyrids (weakly electric fishes), which have unusually elaborate and sophisticated cerebellum, all fishes have small cerebellum. Ancient supra-mammals (ancestral amniotes) show an expansion of cerebellum~\cite{ButlerBook2005}. In accordance, we observe Cerebellum over-represented again in Eutheria. Curiously, one of the genes responsible for the latter is a Purkinje cell protein. Purkinje cells are found in all jawed vertebrates, however, their morphology is remarkably similar in non-amphibian tetrapods~\cite{Smith1993, ButlerBook2005}. It is not entirely clear what the functional role of the genes are that cause over-expression of cerebellum in Eutheria. 

The olfactory areas are over-represented for genes dated to Vertebrates. Studies in evolution of olfaction has focused on the Olfacory Receptor (OR) genes~\cite{Niimura2009, Niimura2009a, CollinBook2003}(these are \emph{not} genes expressed in the brain however). Such studies reveal that olfactory system can be traced back to all chordates, Tetrapods show a dramatic expansion of such genes and the olfactory areas is fishes are underdeveloped, arguing for terrestrial adaption imposing severe selective pressure on olfaction~\cite{Freitag1998, Freitag1999}. Interestingly, Sarcopterygii (lobe-finned fishes) which include Tetrapods as a major clade, show over-expression in Olfactory areas in our analysis. 

Genes related to ion channels like chloride, sodium/calcium exchange channels etc. are responsible for over-representation in Metazoa. The highly-conserved nature of key ion channels across Metazoa has been discussed in the literature~\cite{Jan1992, Anderson2001}. It serves as a test of our analysis that cytoskeletal, signal transduction, ion-channel proteins and amino-acid transporter~\cite{Boudko2005} are the only ones that are over-represented in Metazoa. 

The hypothalamus is not over-represented until genes dated to Mammalia. This is rather curious because hypothalamus is present across vertebrates. The mammalian hypothalamus is highly developed compared to other amniotes, controlling temperature regulation, social and parental behavior, fluid balance, milk flow, reproductory functions, pacemaker for biological rhythm etc. Though several of these behavior are well-developed in birds and reptiles, the mammalian genes responsible for over-representation are pregnancy-related and implicated in circadian-control of mammalian locomotion (cardiotrophin-cytokine (CLC) related  genes)~\cite{Kraves2006}. The hypothalamus is over-represented in Euarchontoglires and Rodentia, alluding to its further sophistication and specialization in primates and rodents.

The midbrain region is represented as early as in Metazoa and reappears in Euarchontoglires. The optic tectum, an important part of midbrain, is one of the most conserved structure in the brains of vertebrates~\cite{ButlerBook2005, StriedterBook2005}. However, our results seem to imply that major evolutionary changes has occurred in the Euarchontoglires clade. The thalamus is also present in all vertebrates. We see it appear in Chordates, perhaps owing to the differential elaboration of sensory and motor functions in early chordates/craniates~\cite{CollinBook2003}. Note that glial fibrillary acidic protein (cytoskeletal protein of astroglia) is one of the genes responsible for over-expression of regions in chordates---glial cells are present in almost all vertebrates~\cite{ButlerBook2005}. 

The dorsal stiatopallidal complex has been identified in most branches of vertebrates where its GABAergic population has been reported~\cite{ButlerBook2005}, major changes have occurred in amniotes. In birds and mammals new pathways have evolved concerning the regulation and initiation of voluntary movements, selected by the challenges of highly complex terrestrial environment. The evolution of the striatopallidal complex is less understood in an-amniotes. In our study, the Pallidum is over-represented in Euarchotoglires, striatum in Rodentia and Mouse-specific genes, suggesting very high degree of specialization for higher mammals. The genes responsible for such over-representation are GABA-receptor related and pituitary-related. The weak representation for cortical regions is striking for all clades, though it is weakly represented in Rodentia in the `fine' parcellation, and the result is open to interpretation. 
 
In the current work, we have attempted to draw conclusions on evolution of the mouse brain using molecular evolution of gene expressed, drawing statistical significance using a large set of genes. A quesiton may remain as to why the molecular evolution of a set of brain-specific genes should leaves a signature on the evolution of neuroanatomical regions. In a complimentary work by two of the authors in this study, it was observed that the correlations in the the gene-expression profile in AGEA can be reproduced with high confidence by assimilating two distinct datasets---the known distribution of several neuronal types and the gene expression profiles of each of these neuronal type~\cite{Grange2011}. The evolution of neuronal types is perhaps a connection between the specificity of neuroanatomical regions in the over-expression we report, though that is not the only interpretation possible. We have offered both genetic and traditional comparative neuroanatomy arguments to connect our observations to existing knowledge, however, the information provided by our approach is quantitative and  complementary. We hope to further our studies in comparative brain evolution using gene-expression (as opposed to anatomical homology) of multiple species, as AGEA-like atlases for other species become available.

\bibliography{ortholog} 

\end{document}